\documentclass[prl,twocolumn,superscriptaddress,nobibnotes,amsmath,amssymb,showpacs]{revtex4}
\usepackage{graphicx}
\usepackage{color}
\usepackage[utf8]{inputenc}

\begin{document}

\title{
Break-up of excited $^{12}$C in three equal-energy $\alpha$
particles
}

\author{Ad. R. Raduta}
\affiliation{Institut de Physique Nucl\'eaire, CNRS/IN2P3,
Universit\'e Paris-Sud 11, Orsay, France}
\affiliation{National Institute for Physics and Nuclear Engineering,
Bucharest-Magurele, Romania}
\author{B.~Borderie}
\affiliation{Institut de Physique Nucl\'eaire, CNRS/IN2P3,
Universit\'e Paris-Sud 11, Orsay, France}
\author{E. Geraci}
\affiliation{INFN, Sezione di Catania, Italy}
\affiliation{Dipartimento di Fisica e Astronomia,
Universit\`a di Catania, Italy}
\affiliation{INFN, Sezione di Bologna and Dipartimento di Fisica,
Universit\`a di Bologna, Italy}
\author{N. Le Neindre}
\affiliation{Institut de Physique Nucl\'eaire, CNRS/IN2P3,
Universit\'e Paris-Sud 11, Orsay, France}
\affiliation{LPC, CNRS/IN2P3, Ensicaen, Universit\'{e} de Caen, 
Caen, France}
\author{P.~Napolitani}
\affiliation{Institut de Physique Nucl\'eaire, CNRS/IN2P3,
Universit\'e Paris-Sud 11, Orsay, France}
\author{M.~F.~Rivet}
\affiliation{Institut de Physique Nucl\'eaire, CNRS/IN2P3,
Universit\'e Paris-Sud 11, Orsay, France}
\author{R.~Alba}
\affiliation{INFN, Laboratori Nazionali del Sud, Italy}
\author{F. Amorini}
\affiliation{INFN, Laboratori Nazionali del Sud, Italy}
\author{G. Cardella}
\affiliation{INFN, Sezione di Catania, Italy}
\author{M. Chatterjee}
\affiliation{Saha Institute of Nuclear Physics,
Kolkata, India}
\author{E. De Filippo}
\affiliation{INFN, Sezione di Catania, Italy}
\author{D.~Guinet}
\affiliation{Institut de Physique Nucl\'eaire, CNRS/IN2P3,
Universit\'e Claude Bernard Lyon 1, Villeurbanne, France}
\author{P.~Lautesse}
\affiliation{Institut de Physique Nucl\'eaire, CNRS/IN2P3,
Universit\'e Claude Bernard Lyon 1, Villeurbanne, France}
\author{E. La Guidara}
\affiliation{INFN, Sezione di Catania, Italy}
\affiliation{CSFNSM, Catania, Italy}
\author{G. Lanzalone}
\affiliation{INFN, Laboratori Nazionali del Sud, Italy}
\affiliation{Universit\`a di Enna ``Kore'', Enna, Italy}
\author{G. Lanzano}
\altaffiliation{deceased}
\affiliation{INFN, Sezione di Catania, Italy}
\author{I. Lombardo}
\affiliation{INFN, Laboratori Nazionali del Sud, Italy}
\affiliation{Dipartimento di
Fisica e Astronomia, Universit\`a di Catania, Italy}
\author{O.~Lopez}
\affiliation{LPC, CNRS/IN2P3, Ensicaen, Universit\'{e} de Caen, 
Caen, France}
\author{C.~Maiolino}
\affiliation{INFN, Laboratori Nazionali del Sud, Italy}
\author{A.~Pagano}
\affiliation{INFN, Sezione di Catania, Italy}
\author{S. Pirrone}
\affiliation{INFN, Sezione di Catania, Italy}
\author{G. Politi}
\affiliation{INFN, Sezione di Catania, Italy}
\affiliation{Dipartimento di Fisica e Astronomia,
Universit\`a di Catania, Italy}
\author{F. Porto}
\affiliation{INFN, Laboratori Nazionali del Sud, Italy}
\affiliation{Dipartimento di
Fisica e Astronomia, Universit\`a di Catania, Italy}
\author{F. Rizzo}
\affiliation{INFN, Laboratori Nazionali del Sud, Italy}
\affiliation{Dipartimento di
Fisica e Astronomia, Universit\`a di Catania, Italy}
\author{P. Russotto}
\affiliation{INFN, Laboratori Nazionali del Sud, Italy}
\affiliation{Dipartimento di
Fisica e Astronomia, Universit\`a di Catania, Italy}
\author{J.P.~Wieleczko}
\affiliation{GANIL, (DSM-CEA/CNRS/IN2P3), Caen, France}

\begin{abstract}
The fragmentation of quasi-projectiles from the nuclear reaction
$^{40}Ca$+$^{12}C$ at 25 MeV/nucleon was used to produce excited states
candidates to $\alpha$-particle condensation. The methodology relies on
high granularity 4$\pi$ detection coupled to correlation function techniques.
Under the assumption that
the equality among the kinetic energies of the
emitted $\alpha$-particles 
and the emission simultaneity
constitutes a reliable fingerprint of $\alpha$ condensation,
we identify several tens of events corresponding to the deexcitation of the
Hoyle state of $^{12}$C which fulfill the condition.
\end{abstract}


\today

\maketitle


Bose-Einstein condensation is known to occur in weakly and strongly
interacting systems such as dilute gases and liquid
$^4$He~\cite{BEC}.
Since more than ten years it is also theoretically shown that for symmetric nuclear matter,
below a critical density, $\alpha$-particle condensation is
favored\cite{ropke_nm_prl1998,beyer_plb2000,sogo_prc2009}. It is expected
to occur at densities smaller than a fifth of the nuclear
saturation density. At higher densities the 2-nucleon deuteron condensation
prevails over the 4-nucleon condensation.
This new possible phase of nuclear matter may have its counterpart in
low-density states of self conjugate lighter nuclei, in the same way
as superfluid nuclei are the finite-size counterpart of superfluid nuclear and
neutron matter.
 This means that under some circumstances,
the alpha condensation, i.e. bosonic properties, might dominate over the nucleon
properties even in finite nuclei.
Thus, by showing that the Hoyle state (i.e. the first excited state 0$^+$ at
7.654 MeV of $^{12}$C) and the sixth  0$^+$ state at 15.097 MeV of $^{16}$O are
described by $\alpha$-particle condensate type functions, 
Refs. \cite{tohsaki_prl2001,funaki_prl2008}
advance the idea that these states are candidates to observe
$\alpha$ condensation. A common feature of these states is their diluteness.
For instance, calculations \cite{tohsaki_prl2001} 
and recent experimental data \cite{ohkubo_prc2004} show that
the rms radius of the Hoyle state exceeds by 45\% the
radius of $^{12}$C in its ground-state.
On the other hand the Hoyle state is known to play a decisive role
in stellar nucleosynthesis of $^{12}$C~\cite{Hoy53,Coo57,Fre94}  with the
assumption that the only contribution to the alpha-particle width comes from
the $^8$Be$_{g.s.}$ + $\alpha$ reaction~\cite{Ueg79}.
At present an experimental upper limit of 4\% exists for the
contribution from the direct 3$\alpha$ channel to the alpha
width \cite{Fre94}.
A phase space calculation suggests that the probability 
ratio of the 3$\alpha$ to the $\alpha$+$^8$Be decay process is 
5 $\times 10^{-4}$ \cite{Fre94}.
However such a factor may be enhanced
in relation with the peculiar properties
of the Hoyle state previously discussed.

According to the present understanding, the
compatibility of a nuclear state with $\alpha$-particle condensation
may be judged upon its excitation
energy close to the $N \alpha$ threshold, the emission simultaneity and
both the low kinetic energy and kinetic energy dispersion. Then, it becomes clear that
probably the most appropriate methodology should involve high velocity
reaction products in the laboratory detected by a
high granularity-high solid angle
particle array, as needed in
multifragmentation reaction studies\cite{desouza_epja2006}.

The Letter reports on results obtained for the nuclear reaction 
$^{40}$Ca+$^{12}$C at 25 MeV per nucleon incident energy performed at INFN,
Laboratori Nazionali del Sud in Catania, Italy. 
The beam impinging on a thin carbon target (320$\mu g/cm^2$)
was delivered by the Superconducting Cyclotron and the
charged reaction products were detected by the CHIMERA 4$\pi$
multidetector \cite{chimera}. The beam intensity was kept around
$10^7$ ions/s to avoid pile-up events.
CHIMERA consists of 1192 silicon-CsI(Tl) telescopes mounted on 35
rings covering 94\% of the solid angle, with polar angle ranging from
1$^{\circ}$ to 176$^{\circ}$. 
Among its most interesting characteristics, 
we mention the low detection and identification thresholds
for light charged particles (LCP)
and the very high granularity at forward angles.  
The mass and charge of the detected
nuclei were determined by standard time of flight (TOF) for
LCP stopped in silicon detectors and
$\Delta E-E$ ($Z>5$) and
shape identification ($Z \leq 5$) techniques
for charged products stopped in CsI(Tl).
In particular $^8$Be nuclei (two equal-energy $\alpha$s 
hitting the same crystal) were identified in CsI(Tl) \cite{Mor10}.
The energy of detected nuclei was measured by the Si detectors
calibrated using proton, carbon and oxygen beams
at various energies ranging from 10 to 100 MeV. 
For $Z=2$, dedicated energy calibrations of the fast component of CsI(Tl)
light was realized using the time of flight (TOF).
Though high-quality TOF charts allowing for a direct calibration of the
CsI light
exist only for 60\% of the total number of modules,
we have finally managed to calibrate more than 95\% of modules
from 1$^{\circ}$ to 62$^{\circ}$.
The modules for which TOF information was poor or missing were calibrated
by comparing the fast component distribution 
with the benchmark distribution of the corresponding ring built out 
the telescopes with excellent TOF.
\begin{figure}
\begin{center}
\includegraphics[angle=0, width=0.95\columnwidth]{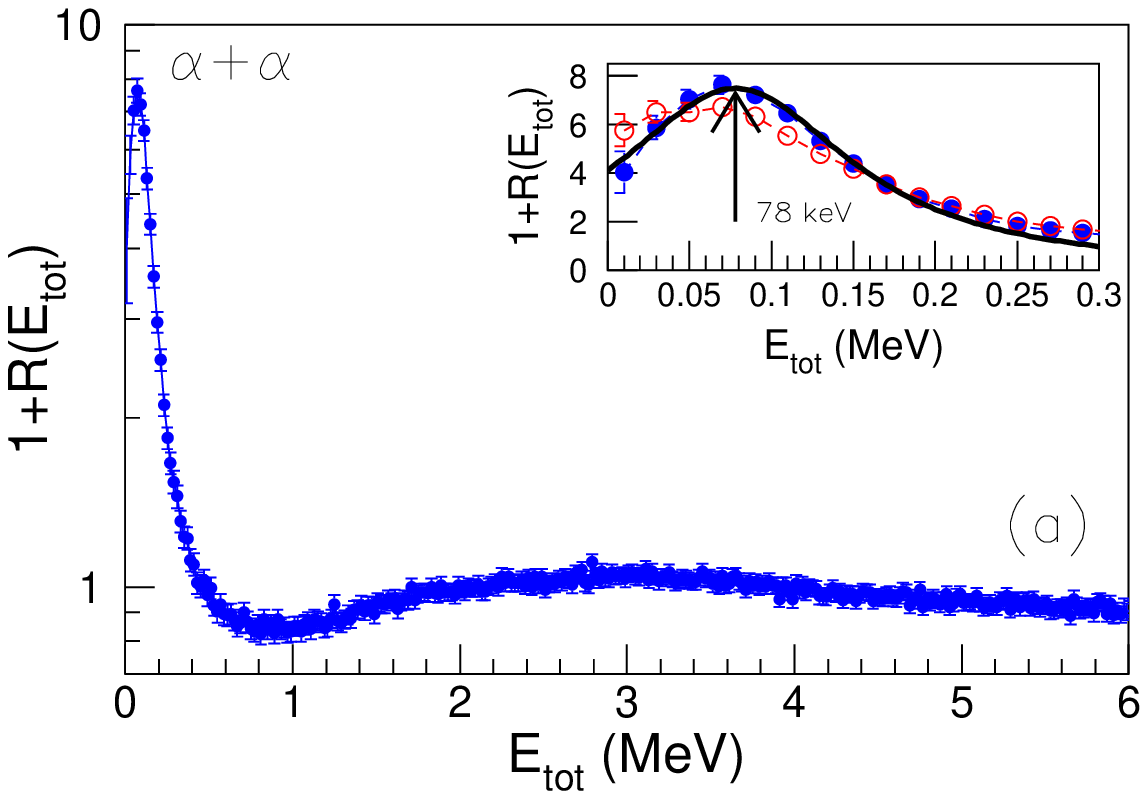}
\includegraphics[angle=0, width=0.95\columnwidth]{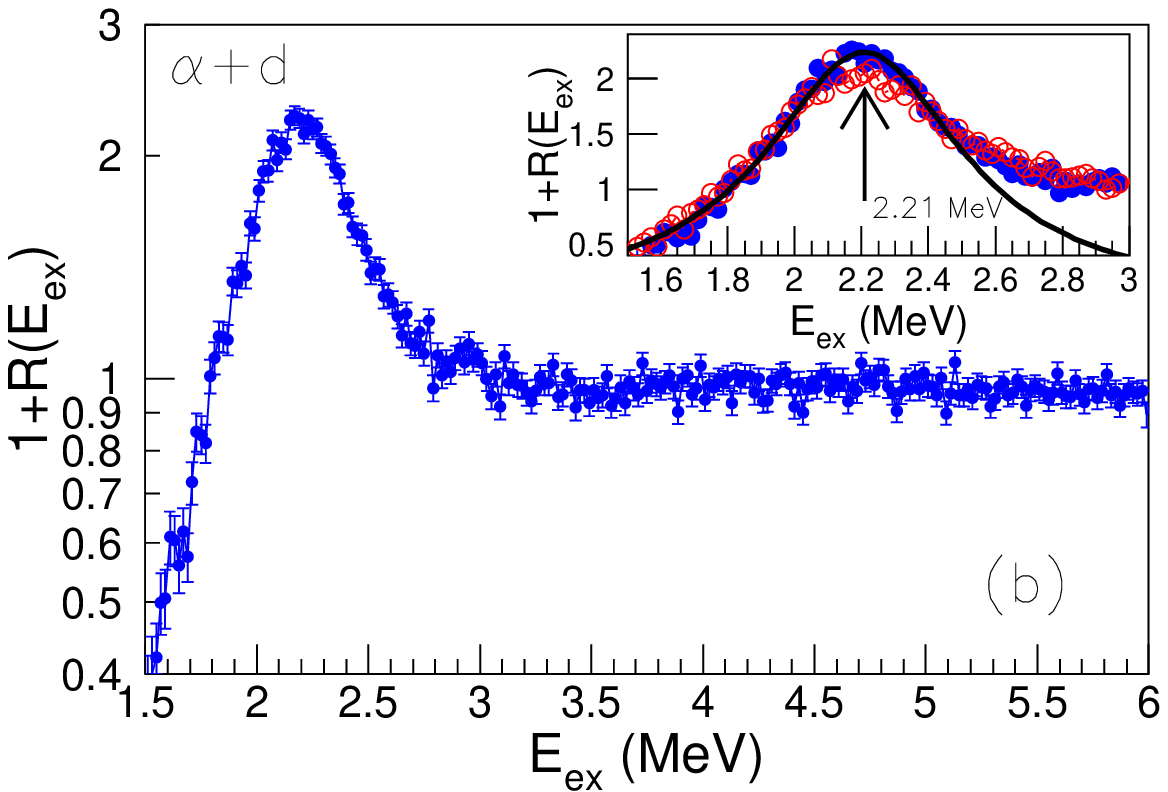}
\end{center}
\caption{(Color online)
Correlation functions in total kinetic energy for $\alpha-\alpha$ (top panel) 
and in excitation energy 
for $\alpha$-d (bottom panel) emission.
Full and open symbols correspond to different ways of calculating the angle
under which the detected particles were emitted (see text). 
The arrows in the insets point the corresponding centroids of the 
Breit-Wigner distributions (solid lines)
fitted on the peaks.  
}
\label{fig:test_calib}
\end{figure}

Invariant $v_{par}-v_{per}$ plots show the binary character of
the collisions with the formation of 
two emitting sources, a quasi-target (QT) and a quasi-projectile (QP). 
These latter may be easily separated according to the smaller or larger
velocity of reaction products with respect to $v_{proj}/2$. 
After a first event filtering according to the total $\alpha$ multiplicity
($m_{\alpha} \geq 3$)
and the total detected charge ($Z_{tot} \leq Z_P+Z_T=26$),
we have focused exclusively on QP decay products with $m_{\alpha} \geq 3$.

The power of correlation functions was extensively discussed in the
literature~\cite{Kim92,charity_prc1995,verde_epja2006,gabi_epja2003}.
They can furnish space-time information taking advantage of
proximity effects induced by Coulomb repulsion, reveal excited
states (whose decay produces the detected correlated particles) and
enlighten any production of events or subevents with specific partitions.
In this Letter various correlation functions have
been used: two-particle correlation functions to first judge the quality
of energy calibrations and multi-particle correlation functions to identify
possible $\alpha$-particle condensation states and define their de-excitation
characteristics.

Two- or multiple-particle correlation functions, with only one variable
E are defined as:
\begin{equation}
1+R(E)=\frac{Y_{corr}(E)}{Y_{uncorr}(E)},
\label{eq:corrf}
\end{equation}
where the role of the generic variable $E$ is equivalently played by 
the total kinetic energy of the particles of interest in their center-of-mass
frame $E_{tot}$ or by the excitation energy of their emitting source/state, 
$E_{ex}=E_{tot}-Q$.
In Eq. (\ref{eq:corrf}) $Y_{corr}$, the correlated yield spectrum, is
constructed with the considered particles in the same event
and $Y_{uncorr}$ stands for the uncorrelated yield spectrum constructed
by taking particles in different events.
In addition to these, by comparing the width of the peaks of the correlation function 
with the natural width of the excited states, one can estimate the 
distortions caused by the finite granularity.
Fig. \ref{fig:test_calib} presents the two-particle correlation functions for 
$\alpha-\alpha$ (a) and $\alpha$-d (b)
in terms of total kinetic energy and, respectively, excitation energy.
The error bars have been calculated taking into account the statistical errors
on both the correlated and uncorrelated spectra.
For $\alpha-\alpha$ we considered the QP event sample described in the paragraph above. 
For $\alpha$-d, we have additionally set the deuteron multiplicity to at
least 1. 
One may see that the $\alpha-\alpha$ correlation function shows a 
narrow peak centered at 78 keV ($\Gamma$=88 keV) which corresponds to the
ground state of $^8$Be ($Q$=-92 keV) and a much broader peak centered at
3.17 MeV ($\Gamma$=4.80 MeV), corresponding to the first excited state at 
3.03 MeV. 
For $\alpha$-d we get a well formed peak centered at the excitation energy
2.21 MeV ($\Gamma$=0.36 MeV)
which definitely corresponds to the first excited state of 
$^6$Li at 2.186 MeV. 
The broadening of correlation peaks is the genuine consequence of detector
finite granularity. 
The finite granularity is responsible also for a certain imprecision in the
determination of the energy.
To illustrate this, Fig. \ref{fig:test_calib} presents the correlation
functions obtained under different hypotheses on the angle under which the
detected particles were emitted. In the first case (solid symbols) we considered
that all the particles which hit a certain module have been emitted under the
angle corresponding to the geometrical center of that module. An alternative
solution is to attribute to each
particle a random angle in the domains allowed
by the geometrical extension of the detector (open symbols). As we have
noticed that the first hypothesis leads to smoother distributions and does
not introduce extra broadening, the fits are done on the corresponding curves. 
This first hypothesis will be used for what follows.
Given the complexity of the apparatus, 
one may conclude that the energy calibration is excellent and
proceed to further spectroscopic analyses.
 
\begin{figure}
\begin{center}
\includegraphics[angle=0, width=0.95\columnwidth]{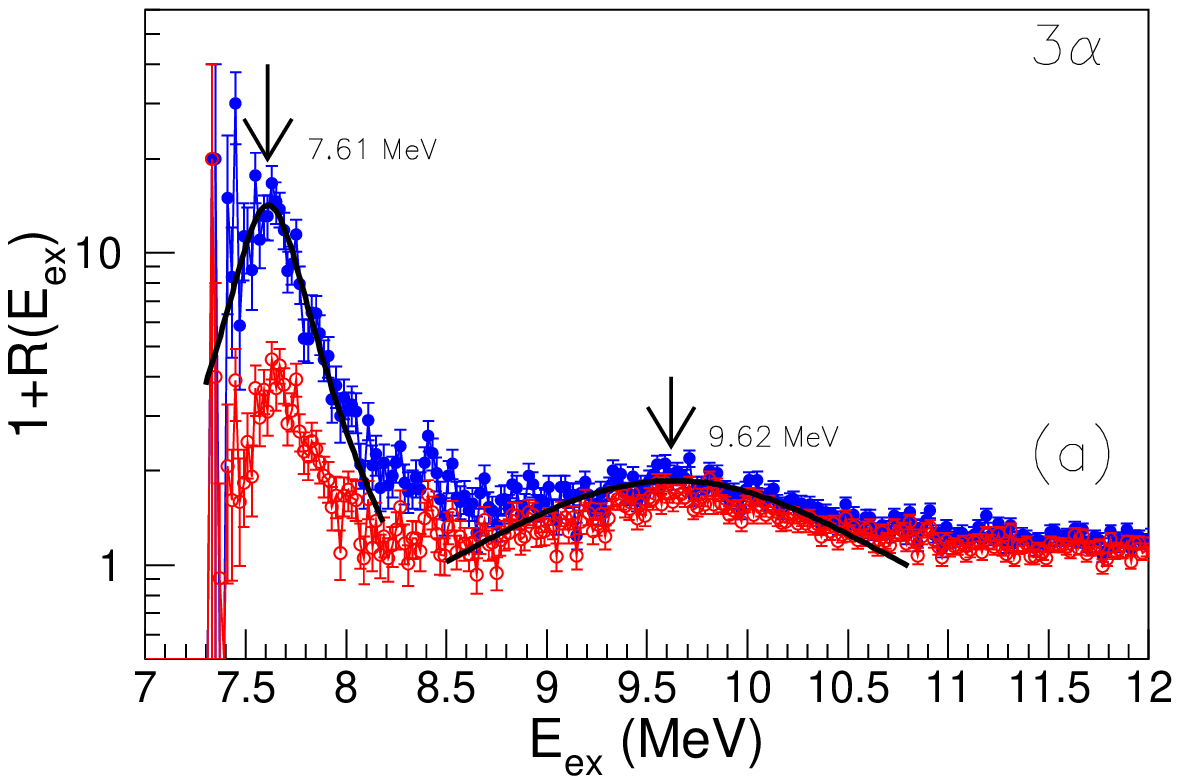}
\includegraphics[angle=0, width=0.95\columnwidth]{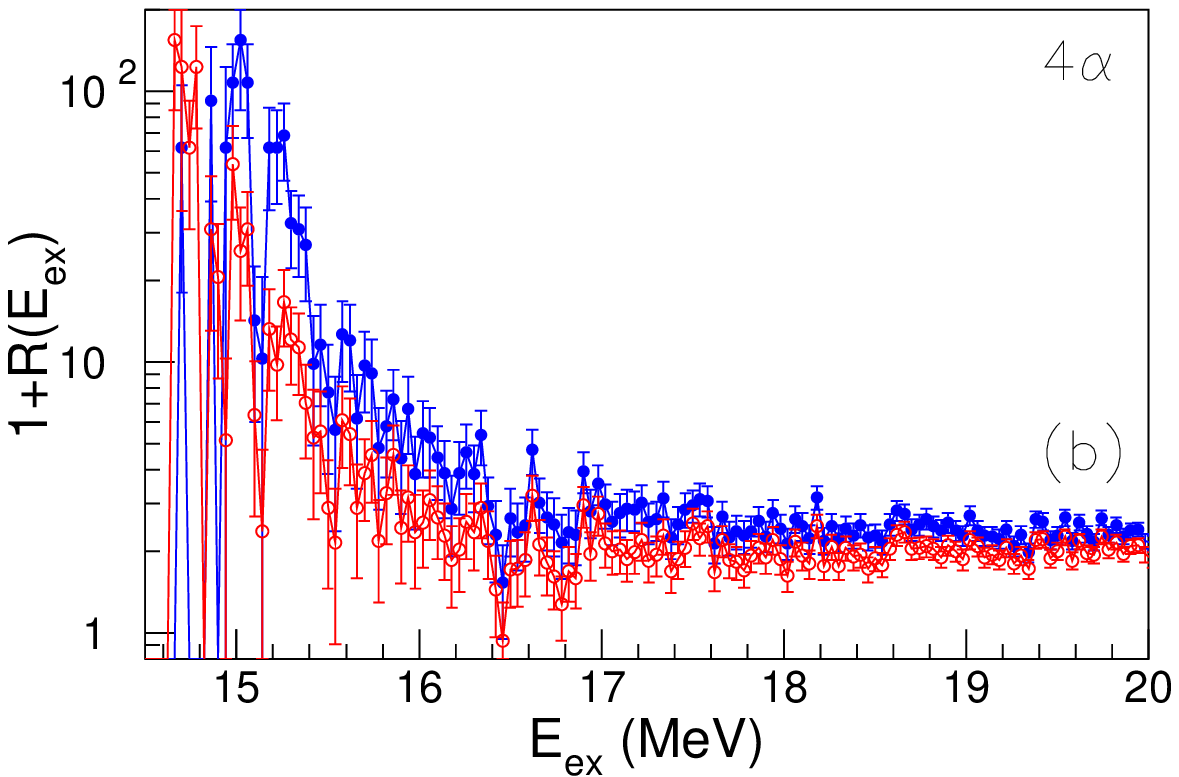}
\end{center}
\caption{(Color online)
Correlation functions in excitation energy for 3-$\alpha$ (top) and
4-$\alpha$ (bottom) emission. 
The distributions plotted with solid circles correspond to the case in which
the uncorrelated yield is built by considering that each particle comes from a
different event.
The distributions plotted with open circles correspond to the case in which
the uncorrelated yield is built by considering emission through 
$^8$Be, $^{12}$C, 2 $\times$ $^8$Be intermediate states.
The arrows correspond to centroids of Breit-Wigner distributions (solid lines)
}
\label{fig:corrf}
\end{figure}

To do this, we exploit multi-particle correlation functions
as shown in Fig. \ref{fig:corrf} for 3-$\alpha$ (a) 
and 4-$\alpha$ (b) emissions.
In the first case we consider the QP events with $m_{\alpha}=3$ and
in the second case (poorer statistics) we
consider QP events with $m_{\alpha} \geq 4$.
The error bars are calculated
considering only the statistical errors of the correlated spectra.
Indeed errors on the uncorrelated spectra have been reduced to negligible
values by increasing the number of uncorrelated events as
compared to correlated events. 
The 3-$\alpha$ correlation function 
with standard event mixing for building the
uncorrelated yield spectrum (i.e. each particle belongs to a different event,
solid symbols)
shows two peaks, at $E_{ex}$=7.61 MeV ($\Gamma$=0.18 MeV) and 
9.62 MeV ($\Gamma$=1.33 MeV). They correspond respectively to 
the Hoyle state ($E_{ex}^{exp}$=7.654 MeV, $\Gamma^{exp}$=8.5 eV)
and to the complex excited region of $^{12}$C,
characterized by the strong $E_{ex}^{exp}$=9.64 MeV, $3^-$ state and
by the broad $E_{ex}^{exp}$=10.3 MeV, $0^+$ 
state submerging a possible $2^+$ state at 9.7 MeV~\cite{itoh_npa2004,freer_npa2010}.
To have a first indication on the extent to which these states may be
considered candidates for $\alpha$-particle condensation,
it is mandatory to check whether the decay is
simultaneous or if it proceeds partially via $^8$Be.
To do this, we adopt the procedure proposed in Ref. \cite{grenier_npa2008} and 
plot with open symbols the correlation function obtained with partial event mixing 
(i.e. the uncorrelated yield spectrum is built by selecting two $\alpha$ particles
from the same event and the third one from a different event). 
As one may see, the two peaks survive, but while the peak at higher
excitation energy is almost
unchanged, the magnitude of the Hoyle state is diminished.
This is a qualitative indication that decays proceeding via
$^8$Be play a different role in the two cases.
The bottom panel of Fig. \ref{fig:corrf} shows that the spectroscopic
information one may extract out of the 4-$\alpha$ correlation functions is much
poorer. The first explanation for the modest quality of those correlation
functions is the reduced statistics generating large error bars. 
But an even more important and fundamental reason
is that in this energy domain the density of
states which decay via $\alpha$ emission is high and, not less
important, most of them are rather broad. 
Indeed, a numerical simulation where simultaneous 4-$\alpha$ decay is assumed
to occur with a branching ratio of 1
from each excited state with completely missing decay information or
which is known to decay via $\alpha$-emission
shows that, after filtering by CHIMERA,
the only state above the 4-$\alpha$ emission threshold
which may be identified is the one at 15.78 MeV.
The solid symbols in Fig. \ref{fig:corrf} correspond to standard event
mixing for the uncorrelated yield spectrum. Open symbols correspond to
partial event mixing considering decays
via intermediate $^8$Be, $^{12}$C and 2 $\times$ $^8$Be. 
The combined background considers the relative probability of these decays.
In the region of interest, around $E_{ex}$=15.1 MeV, a lower reduction
(as compared to the one of the Hoyle state) is
observed between the two functions. This seems to be an indication that the direct 
4$\alpha$ decay is more favourable due to larger decay barriers involved with
$^8$Be and $^{12}$C decays~\cite{freer_mpe2008}.

\begin{figure}
\begin{center}
\includegraphics[angle=0, width=0.95\columnwidth]{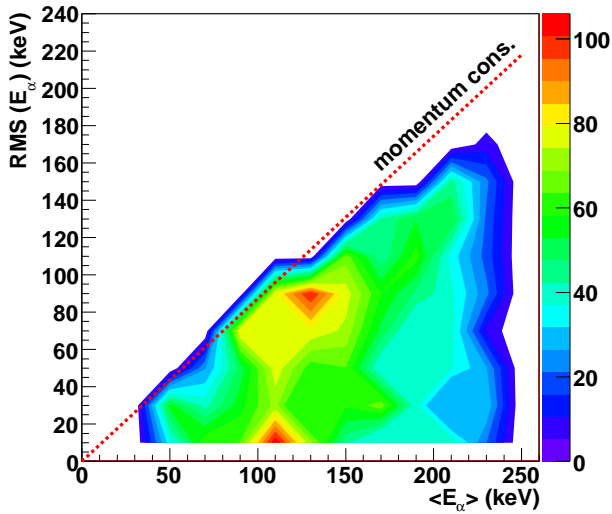}
\end{center}
\caption{(Color online)
Three-$\alpha$ intra-event correlation function expressed as
a function of average kinetic energy - $\sigma_{E_{\alpha}}$ (RMS)
of $\alpha$ particles.
The uncorrelated yield is built such to allow for decay through $^8$Be.
The dotted line marks the maximum RMS compatible with momentum
conservation.}
\label{fig:iev_12c}
\end{figure}
To summarize, we demonstrated so far that the $^{40}$Ca+$^{12}$C 
nuclear reaction at 25 MeV/nucleon
populates excited states of both $^{12}$C and $^{16}$O
nuclei which decay by 3- and, respectively, 4-$\alpha$
emission.
Certainly this part of information is not sufficient to conclude that those
states represent $\alpha$-particle condensation. A major step in this
direction is to show that the emitted
particles have practically the same kinetic energy.
To do that an intra-event correlation method is used. 
The correlation function reads:
\begin{equation}
1+R(\sigma_{E_{\alpha}},<E_{\alpha}>)=
\frac{Y_{corr}(\sigma_{E_{\alpha}}, <E_{\alpha}>)}
{Y_{uncorr}(\sigma_{E_{\alpha}}, <E_{\alpha}>)}.
\label{eq:incorrf}
\end{equation}
Here, for a given alpha multiplicity, the numerator is the yield of
events with given average kinetic energy of alpha particles,
$<E_{\alpha}>$, and given root mean square, $\sigma_{E_{\alpha}}$.

Fig.\ref{fig:iev_12c} illustrates the intra-event correlation function
corresponding to 3-$\alpha$ decays from the region of the Hoyle state
(7.375 MeV$\leqq E_{ex} \leqq$7.975 MeV); the denominator is built
using partial event mixing.
Around the energy of the Hoyle state above the 3-$\alpha$ threshold
($<E_{\alpha}>$= 379/3 keV) two peaks are observed. Qualitatively
the same picture is obtained with standard event mixing
 and also when removing the restriction on the excitation 
energy range.
It is remarkable
to notice the peak localized around
$<E_{\alpha}>$=110 keV and $\sigma_{E_{\alpha}} \leqq$ 25 keV.
Those values correspond, within our energy uncertainties (calibration
and direction of velocity vectors),
to an equal sharing of the available energy of the Hoyle state (379 keV)
among the three $\alpha$ particles. 
The number of events which contribute to the peak is 39. 
So we can consider that at least this number of
condensate states have been
populated in the present experiment. 
The peak localized around
$<E_{\alpha}>$=130 keV and $\sigma_{E_{\alpha}}$ around 90 keV corresponds to
the sharing of the available energy between the two $\alpha$s of $^8$Be
and the remaining $\alpha$ of 191 keV. The number of such events is 85
and must be also considered as a lower limit. Indeed,
depending on their velocity and polar angle in the laboratory, $^8$Be from the
Hoyle state can decay by firing a single telescope. 
For the sake of completeness, we mention that the total
number of events with ($m_{8Be}$=1, $m_{\alpha}=1$) 
in the region of the Hoyle state is 900 while that corresponding to
$m_{\alpha}=3$ is 1072.
Numerical simulations, filtered by the
multi-detector replica, confirm that the observed peaks in
fig.~\ref{fig:iev_12c} correspond to 3$\alpha$ decay (RMS$<$20 keV) and 
$^8$Be-$\alpha$ decay, with the two $\alpha$ flying along the initial 
$^8$Be direction ($<E_{\alpha}> \sim$ 130 keV, RMS $\sim$85 keV). The other
extreme case, $^8$Be-$\alpha$ decay, with the two $\alpha$ flying 
perpendicularly to the initial $^8$Be direction would fall at
($<E_{\alpha}> \sim$ 90 keV, RMS $\sim$60 keV). 
The branching ratio between the
two decay processes, very delicate to estimate, can  not yet be given.
Concerning the 3-$\alpha$ decays of the broad 9.64 MeV state,
we mention that no peak at low $\sigma_{E_{\alpha}}$ was observed in the
intra-event correlation function.
The much poorer statistics we have for 4-$\alpha$ does not allow a similar
analysis. Nevertheless, we mention that there are four events compatible with
the above stated criteria and corresponding to the excited state at 15.1
MeV of $^{16}$O~\cite{ohkubo_plb2010}.

In conclusion, the nuclear reaction $^{40}$Ca+$^{12}$C at 
25 MeV/nucleon bombarding energy was used to produce states 
theoretically predicted as $\alpha$-particle condensate states.
Supposing that equal values of kinetic energy of the emitted
$\alpha$-particles
represent a sufficient criterion for deciding in favor of $\alpha$-particle
condensation,
we found 39 events corresponding to the direct 
de-excitation of the Hoyle state of $^{12}$C
which satisfy the criterion. It also clearly appears that the major part of
Hoyle states deexcite via the $^8$Be$+ \alpha$ channel.
Though the poor statistics prevents drawing a definite conclusion in the 
case of $^{16}$O,
we mention the existence of a small number of events originating from the
decay of the 15.1 MeV state which fulfill the present criteria.
An experiment with higher statistics is planned to better study
the $^{16}$O case.

Acknowledgements.
The authors are indebted to P.~Schuck for numerous
discussions and one of the authors Ad. R. R. acknowledges the partial financial support 
from ANCS, Romania, under grant Idei nr. 267/2007.


\end{document}